# On-Demand Inverse Design for Narrowband Nanophotonic Structures Based on Generative Model and Tandem Network

Yuxiao Li[1], Taeyoon Kim[1], Allen Zhang[2], Zengbo Wang[3,*], Yongmin Liu[1,2,*]

[1]Department of Electrical and Computer Engineering, Northeastern University, Boston, Massachusetts 02115, USA

[2]Department of Mechanical and Industrial Engineering, Northeastern University, Boston, Massachusetts 02115, USA

[3]School of Computer Science and Engineering, Bangor University, Bangor LL57 1UT, UK

*Corresponding authors. Email: z.wang@bangor.ac.uk, y.liu@northeastern.edu

**Abstract**

Inverse design in nanophotonics remains challenging due to its ill-posed nature and sensitivity to input inaccuracies. We present a novel framework that combines a Conditional Variational Autoencoder (CVAE) with a tandem network, enabling robust and efficient on-demand inverse design of photonic structures. Unlike prior approaches that use CVAEs or tandem networks in isolation, our method integrates spectral adjustment and structural prediction in a unified architecture. Specifically, the CVAE adjusts the idealized target spectra, such as Lorentzian-shaped notches, making them more physically realizable and consistent with the training data distribution. This adjusted spectrum is then passed to the tandem network, which predicts the corresponding structural parameters. The framework effectively handles both narrowband (<50 nm) and highly complex spectra, while addressing the one-to-many mapping challenge inherent in inverse design. The model achieves high accuracy, and the designed spectra closely match full-wave simulation results, validating its practicality for advanced nanophotonic applications.

**INTRODUCTION**

Nanophotonics focuses on the study and manipulation of light at the nanometer scale. The design of nanophotonic structures, including optical metamaterials and metasurfaces [1–5], photonic crystals [6,7], as well as plasmonic nanostructures [8,9], has become increasingly complex, posing significant challenges to traditional design methods. The conventional approaches rely heavily on designer expertise, drawing on physical insights from analytical models, and extensive numerical simulations to solve Maxwell's equations. While effective, these processes are typically time-consuming and computationally expensive. Inverse design offers a more efficient alternative by determining structure directly from the target performance. However, commonly used optimization strategies, such as genetic algorithms [10], level set methods [11], and topology optimization [12], often suffer from high computational cost and poor scalability when design complexity increases. With the rapid development of stochastic machine learning techniques, including deep learning [13,14], decision trees [15], and random forests [16], significant breakthroughs have been achieved in diverse fields such as image processing [17,18], speech recognition [19], and remote sensing [20]. This momentum has extended into nanophotonics, empowered by machine learning [21-24]. A wide range of studies has demonstrated the use of machine learning in the design of optical neural networks [25–29], metamaterials and meta-devices [30–36], plasmonic nanostructures [37–40], and passive radiative cooling films[41–43]. These works have shown that machine learning is a powerful tool for efficiently identifying the optimal or near-optimal design, especially when the design space is vast or the requirements are stringent, compared to traditional approaches.

Despite significant interest and progress, inverse design in nanophotonics remains inherently challenging due to the ill-posed nature of the problem—where a given optical spectrum may correspond to multiple valid structures or no physically realizable structure at all [34,44,45]. These difficulties become particularly pronounced when targeting narrowband or highly complex spectra, such as single-notch and double-notch filters with sub-50 nm bandwidths. To address these issues, researchers have developed various machine learning-based frameworks, including tandem networks and conditional generative models. Tandem architecture, initially proposed by Liu *et al.* [34], integrates an inverse network with a forward simulator to enforce spectral consistency and physical plausibility. It has been successfully applied to implement metasurfaces [46–48], beam shaping elements [49], and multilayer films [50]. On the other hand, Conditional Variational Autoencoders (CVAEs) have emerged as a powerful tool for generative modeling in nanophotonics, enabling the generation of valid structures conditioned on spectral targets [35, 41–53]. These models can capture the uncertainty and multimodal nature of inverse mappings, which are prevalent in many-to-one scenarios.

Previous studies use CVAEs and tandem networks independently. In this work, we present a unified framework, for the first time to our knowledge, that combines the spectrum-refining capability of a CVAE with the predictive precision of a tandem network for inverse nanophotonic design. Specifically, the CVAE modifies the user-defined ideal target spectrum—such as a Lorentzian-shaped notch—so that it lies within the distribution of physically realizable spectra learned from training data. This CVAE-adjusted spectrum is then passed to the tandem network, which deterministically predicts the structural parameters that can generate the desired spectrum. By combining these two stages, the model improves both spectral fidelity and inverse mapping stability, enabling robust, on-demand design of photonic structures with narrowband (<50 nm) or highly complex transmission spectra. We propose both an end-to-end and a hybrid model framework for future on-demand inverse design. This approach enhances flexibility and practicality, supporting applications such as photonic devices, metasurfaces, and sensors. It lays the groundwork for more efficient and physically grounded inverse design strategies.

**METHOD**

In inverse design, the goal is to determine the structural parameters that yield a desired optical response. Our proposed framework streamlines this process by integrating a CVAE with a tandem network, as illustrated in Fig. 1. In the following, we will discuss the model structure, inference flow, and loss functions associated with the proposed inverse design framework.

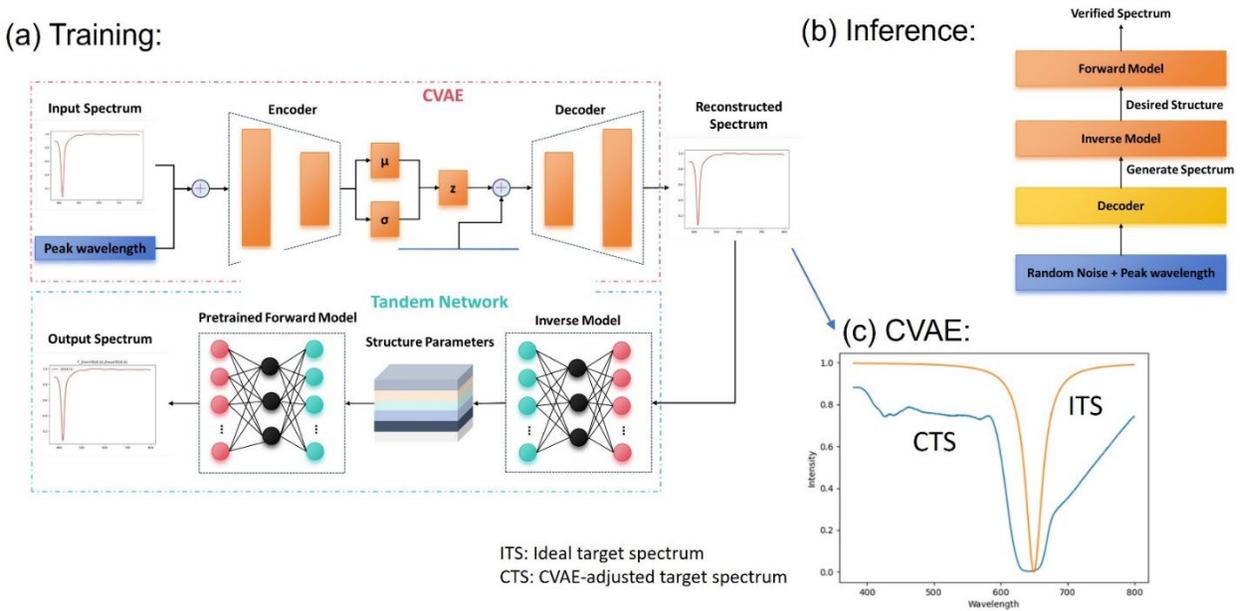

Figure 1. Schematic of the proposed inverse design framework. (a) Model architecture during training. (b) Inference flow based on a user-specified spectrum. (c) Example spectrum transformation from the ideal target spectrum (ITS) to the CVAE-adjusted target spectrum (CTS).

*A. Model Architecture*

The framework comprises two main components: a CVAE for target spectrum refinement and a tandem network for structural parameter prediction. As shown in Fig. 1(a), these modules are trained sequentially.

The CVAE is a generative model conditioned on specific spectral characteristics, such as notch peak wavelength. It encodes the input spectrum into a latent space and then decodes it to reconstruct a spectrum that lies within the physically realizable spectral distribution. Unlike a standard Variational Autoencoder (VAE), the CVAE incorporates conditional information (*e.g.*, the notch peak wavelength) during both encoding and decoding. This allows the model to generate spectra that are not only realistic but also aligned with specific user-defined goals.

During training, the CVAE learns to generate physically realistic spectra—referred to as CVAE-adjusted Target Spectra (CTS)—based on specified physical characteristics, such as target resonance positions or bandwidths. Unlike analytically defined Ideal Target Spectra (ITS), such as Lorentzian notches, which are often oversimplified and may not correspond to physically realizable structures, the CTS incorporate the statistical and physical priors learned from the training data. In this way, the CVAE bridges the gap between the idealized spectral objectives and the actual constraints of nanophotonic device, ensuring that the generated spectrum is both physically consistent and designable.

The tandem network consists of an inverse model and a pre-trained forward model. The forward model is first trained to predict the optical response given structural parameters. The inverse model is then trained to map a given spectrum to a set of structural parameters, which are validated through the forward model. The closed-loop framework ameliorates the intrinsic ill-posedness of the inverse problem and enforces physical consistency in the generated outputs. A comprehensive exposition of the model architecture is presented in Section 1 in the Supplementary Material.

## B. Inference Flow

As illustrated in Fig. 1(b), the inference process begins with a user-specified target, such as a desired notch peak wavelength. Rather than relying on an analytically defined Lorentzian-shaped ITS, which may not correspond to any physically realizable structure, the proposed framework leverages a CVAE to directly generate a CTS. Conditioned on the specified peak wavelength and random noise, the CVAE decoder produces spectra that retains the intended spectral characteristics, such as notch location, while conforming to the manifold of physically realizable spectra learned during training. This approach effectively overcomes the limitations of ITS by ensuring that the generated spectra are both target-consistent and fabrication-feasible.

The resulting CTS is subsequently input into the inverse model of the tandem network, which predicts the structural parameters most likely to produce the desired spectral response. These predicted parameters are then passed through the forward model to reconstruct the corresponding transmission spectrum. As illustrated in Fig. 1(c), this two-stage design pipeline—from CVAE-guided spectrum generation to tandem-based structural inference—ensures that the final output not only adheres to physical constraints but also closely matches the user-specified spectral features.

This inference workflow enables on-demand inverse design using minimal input (*i.e.*, a single target wavelength) and significantly enhances robustness across narrowband and complex spectral regimes. The CVAE plays a pivotal role in bridging the gap between idealized spectral intent and practical realizability, thereby improving both the accuracy and reliability of the overall system.

## C. Loss Functions

Different loss functions are used in each module to guide learning and ensure that both global spectral characteristics and key local features are accurately captured. For the CVAE module, the total loss function is composed of four key components:

(1) a Mean squared error (MSE) reconstruction loss that enforces overall spectral similarity;

(2) a Kullback-Leibler Divergence (KLD) loss that regularizes the latent space;

(3) a notch peak wavelength loss (NPW) that ensures correct spectral localization;

(4) a notch peak intensity loss (NPI) that preserves spectral depth.

These loss terms are jointly optimized to guide the CVAE in generating physically realistic spectra that faithfully represent the user-specified design intent. Each component is defined as follows:

MSE: The MSE reconstruction loss is used to measure the global similarity between the generated spectrum and the target spectrum. It ensures that the generated spectrum retains the overall spectral shape and accurately reproduces the primary features of the target:

$$\mathcal{L}_{\text{recon}} = \frac{1}{N} \sum_{i=1}^{N} (y_i - \hat{y}_i)^2 \tag{1}$$

where $y_i$ and $\hat{y}_i$ represent the target and generated spectral values at the $i$-th point, and $N$ is the number of sampling points in the spectrum.

KLD: To encourage the latent space to follow a smooth and continuous distribution, the KLD loss is introduced, which minimizes the Kullback-Leibler divergence between the learned latent distribution $q(z|x)$ and a standard normal prior $p(z) = \mathcal{N}(0, I)$:

$$\mathcal{L}_{\text{KLD}} = \frac{1}{2} \sum_{j=1}^{d} (\mu_j^2 + \sigma_j^2 - \ln\sigma_j^2 - 1) \tag{2}$$

where $\mu_j$ and $\sigma_j$ denote the mean and standard deviation of the $j$-th dimension in the latent space, and $d$ is the dimension of the latent vector. This regularization prevents the model from collapsing into a deterministic mapping and improves the diversity of the generated spectrum.

NPW: Although the MSE loss ensures global similarity, it often fails to enforce local spectral features, such as peak positions and intensities, which are critical in photonic designs. Therefore, a peak wavelength loss is introduced to ensure that the generated spectrum exhibits a peak at the desired wavelength. This term penalizes the deviation between the generated peak wavelength $\lambda_{\text{gen}}$ and the target peak wavelength $\lambda_{\text{target}}$:

$$\mathcal{L}_{\text{peak-}\lambda} = (\lambda_{\text{gen}} - \lambda_{\text{target}})^2 \tag{3}$$

where $\lambda_{\text{gen}}$ and $\lambda_{\text{target}}$ represent the generated and target peak wavelengths, respectively.

NPI: Similarly, a peak intensity loss is introduced to ensure that the intensity at the peak wavelength matches the target intensity. This loss penalizes the difference between the generated peak intensity $I_{\text{gen}}$ and the target peak intensity $I_{\text{target}}$:

$$\mathcal{L}_{\text{peak-I}} = (I_{\text{gen}} - I_{\text{target}})^2 \tag{4}$$

where $I_{\text{gen}}$ and $I_{\text{target}}$ represent the generated and target peak intensities, respectively.

Finally, the combined CVAE loss is formulated as:

$$\mathcal{L}_{\text{CVAE}} = \alpha_1 \mathcal{L}_{\text{recon}} + \beta_1 \mathcal{L}_{\text{KLD}} + \gamma_1 \mathcal{L}_{\text{peak-}\lambda} + \delta_1 \mathcal{L}_{\text{peak-I}} \tag{5}$$

where $\alpha_1$, $\beta_1$, $\gamma_1$, and $\delta_1$ are hyperparameters that control the relative importance of each loss term. Through this composite loss, the CVAE is guided to generate optical spectra that not only align with the global spectral shape but also precisely matches the key spectral features, such as peak position and intensity, while maintaining variability in the generated samples. This multi-objective optimization scheme is particularly crucial for performance-driven inverse design tasks, where both global and local spectral characteristics are critical for achieving the desired optical performance.

For the tandem network, which operates deterministically, the KLD term is excluded. The corresponding loss function is:

$$\mathcal{L}_{\text{tandem}} = \alpha_2 \mathcal{L}_{\text{recon}} + \gamma_2 \mathcal{L}_{\text{peak-}\lambda} + \delta_2 \mathcal{L}_{\text{peak-I}} \tag{6}$$

where $\mathcal{L}_{\text{recon}}$, $\mathcal{L}_{\text{peak-}\lambda}$, and $\mathcal{L}_{\text{peak-I}}$ are the reconstruction loss, peak wavelength loss, and peak intensity loss, respectively. The weighting coefficients $\alpha_2$, $\gamma_2$, and $\delta_2$ control the relative contributions of each loss term.

## RESULT

In this section, we will introduce the nanophotonic structures of our interest and showcase the training and test results of different sections respectively.

### A. Multilayer Meta-film Structure

The structure designed in this work is a multilayer consisting of three functional regions, as illustrated in Figs. 2(a) and 2(b). The top region is a single layer MgF$_2$ anti-reflection coating (ARC), which reduces reflection and enhances transmission efficiency [54]. Beneath this ARC layer are four functional meta-films, which are responsible for achieving the target single-/double-notch narrowband transmission spectrum. The bottom layer is a glass substrate. This multilayer system is intended to achieve a narrowband transmission spectrum at a specific wavelength, which remains robust under varying incident angles, without exhibiting the commonly observed redshift effect in conventional multilayer designs. For further details of this structure, please refer to Section 2 in the Supplementary Material.

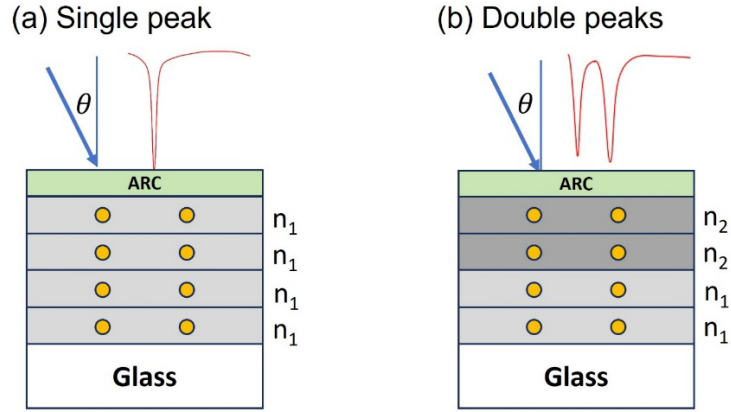

Figure 2. (a) Schematic of the single-notch structure with four identical meta-film layers (refractive index $n_1$) embedded with silver nanoparticles. (b) Schematic of the double-notch structure with alternating meta-film layers (two with refractive index $n_2$ and two with $n_1$). Both structures have an ARC layer on top and a glass substrate at the bottom.

The first dataset is constructed based on multilayer meta-film designed to exhibit a single notch resonance peak. The unit cell of the meta-film consists of a central silver nanoparticle embedded within a multilayer nanostructure. The top $MgF_2$ ARC layer has a fixed height of 99.75 nm and in-plane dimensions of 33.25 nm × 33.25 nm. Each mata-film layer, including the metallic and dielectric layers, shares identical lateral dimensions of 33.25 nm × 33.25 nm and a height of 33.25 nm. The structural parameters are uniform across all layers, ensuring periodicity and consistency throughout the design. The structural variations within this dataset are defined by two key parameters: the radius of silver nanoparticles embedded in the meta-film layers and the refractive index of the meta-film host matrix. The radius of the silver nanoparticles varies from 2.5 to 6.5 nm with a step size of 0.1 nm, while the refractive index of the matrix ranges from 1.5 to 3.05 with a step size of 0.05. These two parameters collectively govern the resonance characteristics of the transmission spectra. Numerical simulations are performed using CST Studio Suite within the visible wavelength range. Periodic boundary conditions are applied to the unit cell to model an infinite array of the multilayer structure.

Considering the constant position of notch peaks at different incidence angles, we averaged the transmission spectra of three different incidence angles to form a new dataset. In total, approximately 1300 samples were generated, covering a comprehensive range of structural variations for the transmission spectra. The second dataset corresponds to structures exhibiting a bimodal narrowband transmission spectrum, characterized by two resonance peaks. To obtain this spectral response, the design variables include the refractive index of the first and second meta-film layers, the refractive index of the third and fourth meta-film layers, and the radius of the embedded silver nanoparticles, which are assumed to be identical across all layers. The remaining structural parameters are consistent with those in the first dataset. In total, approximately 10,000 samples were generated.

## B. Inverse Design for Target Notch Peak Wavelength

In this study, we first conduct training and validation on a unimodal dataset. For further details of train and validation process, please refer to Section 3 in the Supplementary Material.

To evaluate the effectiveness of the CVAE module, we compare the performance of the tandem network when using either ITS or CTS as input. As illustrated in Fig. 3(a), ITS—constructed using the notch peak wavelength at 410 nm, 530 nm, or 615 nm with a bandwidth of 30 nm—is directly fed into the tandem network. The inverse model can predict the refractive index of the meta-film and the radius of the silver nanoparticle, which are subsequently passed into the forward model to reconstruct the transmission spectrum. The output is then compared with the input of the ITS.

It is observed that when the peak wavelength was 410 nm, the tandem network alone is able to reconstruct the transmission spectrum with reasonable accuracy. However, for peak wavelength at 530 nm or 615 nm, the network exhibits significant performance degradation, characterized by evident peak shifts, most notably a 6 nm shift at 530 nm. To address this limitation, we employ a CVAE to generate spectra based on the target peak wavelength and random noise. These spectra are then input into the tandem network, as shown in Fig. 3(b). This hybrid framework, integrating the CVAE and the tandem network, effectively corrects the failure modes observed at 530 nm and 615 nm compared with only tandem network.

Further analysis reveals that when using ITS directly, the tandem network's performance is highly sensitive to bandwidth, and discrepancies between the bandwidth and the physical peak location can lead to substantial errors. In contrast, the CVAE captures the latent representation of transmission spectra consistent with actual structural parameters. Rather than producing ITS, the CTS more faithfully reflects the true physical behavior of the system. As a result, when these CTS are provided as input to the tandem network, the predicted structural parameters are substantially more accurate.

Moreover, the CTS spectra exhibit improved alignment with the training dataset distribution, thereby enhancing the tandem network's robustness and reliability for downstream applications. Finally, the structural parameters obtained through the hybrid approach are validated via full-wave simulations using commercial software CST Studio Suite. The results are presented in Fig. 3(c). The resulting transmission spectra demonstrate very good agreement with those produced by CST Studio Suite, confirming the accuracy of the inverse design. A summary of the inverse design results obtained using only the tandem network versus the CVAE-enhanced framework is provided in Table 1.

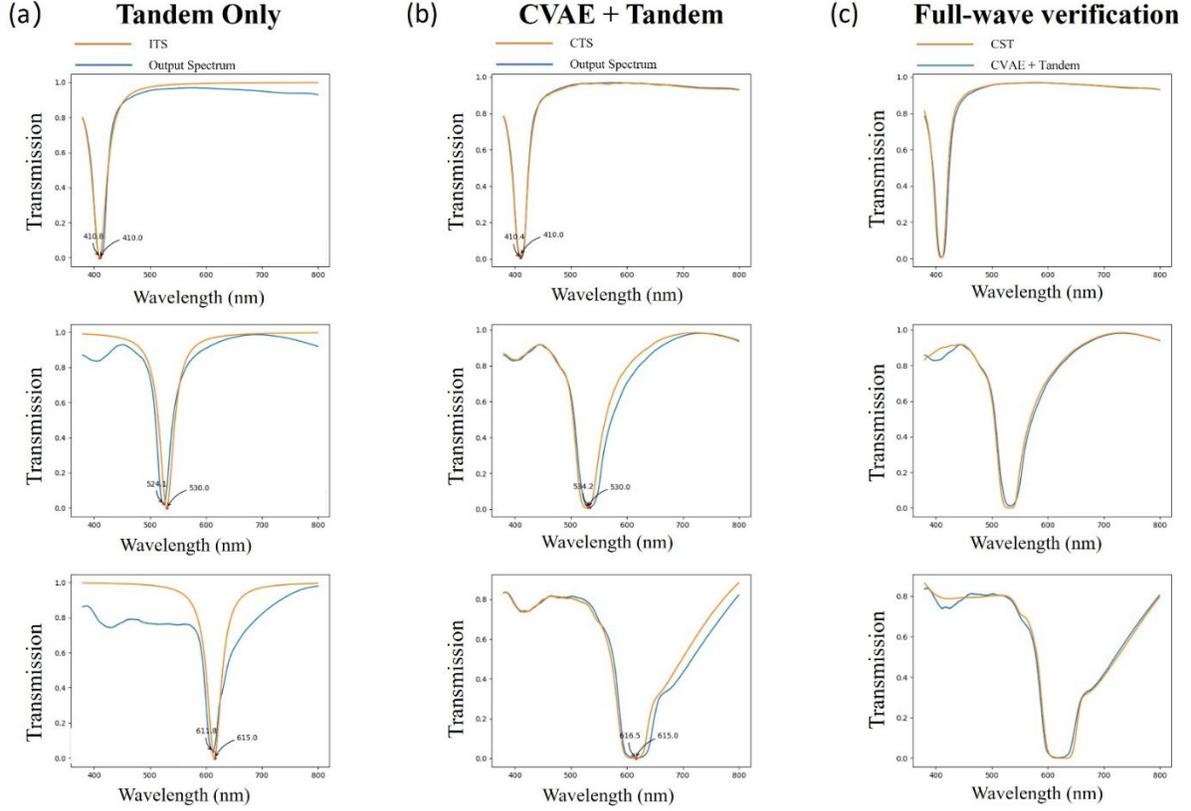

Figure 3. Effect of CVAE-adjusted target spectra on the precise inverse design of physical structures for notch filters at different notch wavelengths. (a) Transmission spectra of the inverse-designed structures based solely on the tandem network with ITS target spectrum. The target notch wavelength is 410, 530, and 615 nm, respectively. (b) Transmission spectra of the inverse-designed structures based on the hybrid CVAE-tandem network with CTS target spectrum. (c) Comparison of the transmission spectra obtained by the hybrid network and CST simulation.

TABLE 1: Single-notch Filter Structure parameters predicted by the tandem-only model and the hybrid CVAE-tandem network.

| Case 1: Single-Notch Filter $|\Delta\lambda| = |\lambda_{a,b} - \lambda_d|$ | | | | | | | | |
|---|---|---|---|---|---|---|---|---|
| **Target Wavelength** $\lambda_d$ **(nm)** | **Tandem Only (ITS)** | | | | **CVAE + Tandem (CTS)** | | | |
| | R | $n_1$ | $\lambda_a$ | $|\Delta\lambda|$ | R | $n_1$ | $\lambda_b$ | $|\Delta\lambda|$ |
| 410 | 5.61 | 1.56 | 410.8 | **0.8** | 5.66 | 1.55 | 410.4 | **0.4** |
| 530 | 4.18 | 2.31 | 524.1 | **5.9** | 5.81 | 2.37 | 534.2 | **4.2** |
| 615 | 3.64 | 2.83 | 611.8 | **3.2** | 6.50 | 2.90 | 616.5 | **0.5** |

For the case of double-notch-peak spectra, we conduct studies on three representative datasets with double-notch wavelengths set to (450 nm, 500 nm), (500 nm, 600 nm), and (430 nm, 620 nm), aiming to compare the performance of the tandem network when using ITS versus CTS as the input. As shown in Fig. 4(a) and 4(b), when ITS are directly fed into the tandem network, the reconstructed transmission spectra exhibit significant peak shifts, particularly at 430 nm and 620 nm. In contrast, when CTS is used as input, these discrepancies are markedly reduced, leading to more accurate spectral predictions. Further comparison between the spectra generated by the CVAE-tandem network and those obtained through CST simulation is shown in Fig. 4(c). The good agreement validates the reliability of the proposed approach. The corresponding structural parameters for each case are summarized in Table 2.

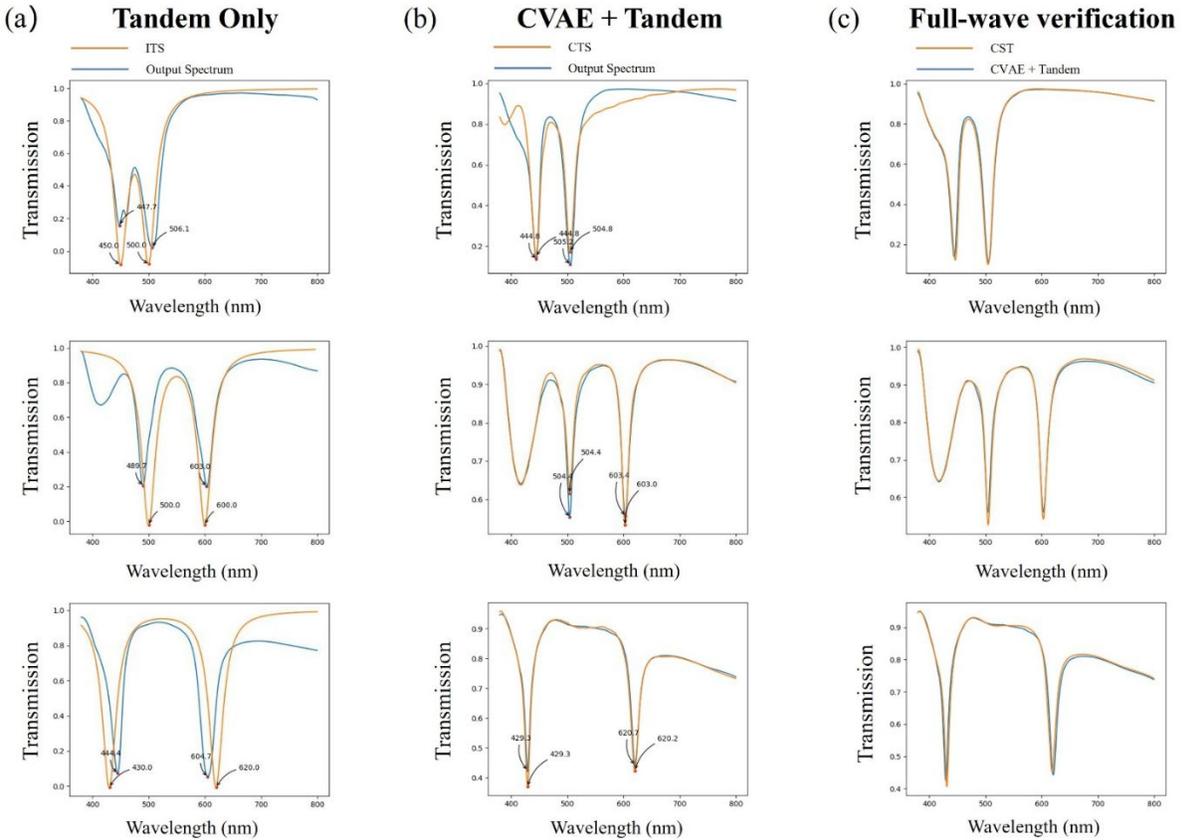

Figure 4. (a) Transmission spectrum of the inverse-designed structures based solely on the tandem network. The double-notch wavelengths are targeted as (450 nm, 500 nm), (500 nm, 600 nm), and (430 nm, 620 nm), respectively. (b) Transmission spectra of the inverse-design structures based on the hybrid CVAE-tandem network. (c) Comparison of the transmission spectra obtained by the hybrid network and CST simulation.

TABLE 2: Double-notch Filter Structure parameters predicted by the tandem model and the hybrid CVAE-tandem network.

| Case 2: Double-Notch Filter $\|\Delta\lambda\| = \|\lambda_{a,b} - \lambda_d\|$ | | | | | | | | |
|---|---|---|---|---|---|---|---|---|
| Target Wavelengths $\lambda_d$ (nm) | Tandem Only (ITS) | | | | CVAE + Tandem (CTS) | | | |
| | R | $n$ ($n_2$, $n_1$) | $\lambda_a$ | $\|\Delta\lambda\|$ | R | $n$ ($n_2$, $n_1$) | $\lambda_b$ | $\|\Delta\lambda\|$ |
| 450 | 5.76 | 2.21 | 447.7 | **2.3** | 4.56 | 2.20 | 444.8 | **5.2** |
| 500 | | 1.85 | 506.1 | **6.1** | | 1.81 | 505.2 | **5.2** |
| 500 | 4.32 | 2.76 | 489.7 | **10.3** | 2.57 | 2.83 | 504.4 | **4.4** |
| 600 | | 2.15 | 603.0 | **3.0** | | 2.17 | 603.0 | **3.0** |
| 430 | 4.85 | 2.79 | 444.4 | **14.4** | 3.26 | 2.89 | 429.3 | **0.7** |
| 620 | | 1.78 | 604.7 | **15.3** | | 1.71 | 620.2 | **0.2** |

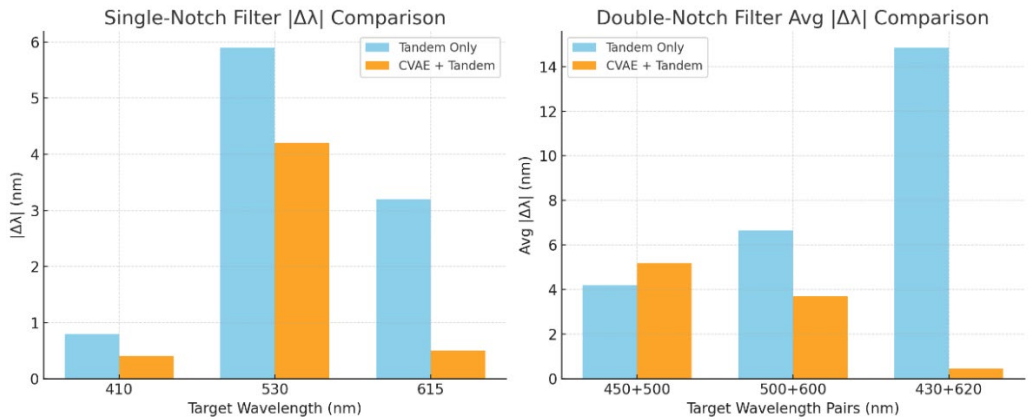

Figure 5. Comparison of wavelength errors ($\|\Delta\lambda\|$) between the Tandem-only and CVAE + Tandem methods for (a) single-notch and (b) double-notch filter designs. The CVAE-based method shows improved accuracy in most cases.

Figure 5 shows a comparison of the wavelength prediction errors ($\|\Delta\lambda\|$) between the Tandem-only (ITS) and the CVAE + Tandem (CTS) methods for both single- and double-notch filter designs. In the single-notch case (Figure 5a), the CTS method achieves consistently lower errors across all target wavelengths. For example, at 410 nm and 615 nm, the error is reduced to 0.4 nm and 0.5 nm respectively, compared to 0.8 nm and 3.2 nm with ITS. This indicates improved precision in capturing the inverse mapping from spectrum to structure.

In the more challenging double-notch cases (Figure 5b), CTS also demonstrates clear advantages. While the ITS method shows large deviations—such as an average error of 14.85 nm for the 430+620 nm target pair—the CTS method reduces this to just 0.45 nm. Similar improvements are observed across other wavelength pairs, suggesting that the CVAE component helps the model generalize better and provide more accurate initial predictions for the tandem refinement stage.

These results confirm that incorporating the CVAE's learned spectral–structural prior improves inverse design—especially for complex multi-peak spectra—by capturing key design patterns in a latent space that stabilizes the mapping and enables more reliable, precise convergence to physically realizable filter parameters.

## DISCUSSION

We demonstrated a compact hybrid inverse design workflow that first "translates" sparse, idealized spectral intentions (one or two target notch wavelengths) into physically realizable CVAE-adjusted spectra (CTS) and then applies a deterministic tandem network to retrieve multilayer meta-film geometries. This front-loaded feasibility transform is the key advance: by regularizing targets onto the empirical spectral manifold, it suppresses the instability and non-uniqueness that undermine direct (tandem-only) inversion of analytic Lorentzian notches.

Consequently, wavelength prediction errors shrink for both single and dual notch filters, with the largest relative gains where conventional inversion previously drifted by several nanometers, and dual-notch peak alignment improves without sacrificing angle robustness. The approach thus unifies generative realism (via the CVAE prior) and forward-model physical consistency (via the tandem stage) in a simple, modular sequence. Overall, conditioning inputs through a learned feasibility transform provides a general template for other ill-posed nanophotonic inverse problems where user-specified analytic targets diverge from device-achievable responses.

## CONCLUSION

In conclusion, we propose a hybrid model that integrates a CVAE with a tandem network to address the challenge of generating physically meaningful input spectra in inverse design. The CVAE generates spectra with realistic physical characteristics based on spectral features, ensuring greater consistency with the spectra produced by the original structures. Compared to arbitrarily assigned spectra, such as ITS, CTS more accurately reflect the intrinsic properties of the original system, thereby reducing the structural error in the inverse design process. This approach offers a potential solution for more flexible and efficient inverse design in the field of nanophotonics.

## ACKNOWLEDGMENTS

Z.W. gratefully acknowledges support from the Leverhulme Trust (RF-2022-659) and joint support from the Leverhulme Trust and the Academies (British Academy, Royal Academy of Engineering, and Royal Society) for grant APX\R1\251114. Additional support was provided by Bangor University (BUIIA-S46910).

# Supplementary Material

## Section 1: structure information of model

Table S1: Definition and parameters of the neural networks. "FC" refers to fully connected layer followed by batch normalization layer and ReLU activation function.

| Neural Network | Layer | Number of neurons |
|---|---|---|
| CVAE (Encoder) | Input Layer | 1002 (comprising 1001 spectrum features and 1 conditional parameter) |
| | FC_1 | 1024 |
| | FC_2 | 2048 |
| | FC_3 | 1024 |
| | FC_4 | 512 |
| | FC_5 | 256 |
| | FC_6 | 128 |
| | Gaussian Mean / Gaussian Covariance | 25 |
| CVAE (Decoder) | Input Layer | 26 (comprising 25 latent variables and 1 conditional parameter) |
| | FC_7 | 128 |
| | FC_8 | 256 |
| | FC_9 | 512 |
| | FC_10 | 1024 |
| | FC_11 | 2048 |
| | FC_12 | 1024 |
| | FC_13 | 1001 |
| Tandem Network (Inverse Model) | FC_14 | 1024 |
| | FC_15 | 2048 |
| | FC_16 | 1024 |
| | FC_17 | 512 |
| | FC_18 | 256 |
| | FC_19 | 128 |
| | FC_20 | 64 |
| | Structure Parameters | 2 (single-notch structure)/ 3 (double-notch structure) |
| Tandem Network (Forward Model) | FC_21 | 64 |
| | FC_22 | 128 |
| | FC_23 | 256 |
| | FC_24 | 512 |
| | FC_25 | 1024 |
| | FC_26 | 2048 |
| | FC_27 | 1024 |
| | Output Layer | 1001 |

The detailed configurations of the proposed hybrid network are presented in Table S1. All networks employ fully connected layers to process the input features. Each layer is followed by batch normalization and a Rectified Linear Unit (ReLU) activation function to enhance training stability and nonlinearity. For the decoders in the CVAE, the inverse model of the tandem network, and the forward model, a Sigmoid activation function is applied at the output layer to constrain the predictions within a physically meaningful range. The final structural parameters are obtained by scaling the normalized outputs with the original boundary values of the design variables, thereby yielding the final design specifications.

**Section2: Multilayer Meta-film Structure**

Fig. S1(a) schematically show of proposed multilayer structure. Unlike traditional multilayer structures that typically rely on naturally available homogeneous materials, here the functional meta-film incorporates an artificial metamaterial. Fig. S1(b) illustrates the unit cell of the meta-film. It consists of a silver nanoparticle embedded in a dielectric host medium. The presence of such a composite structure introduces localized plasmonic resonance and enhances light confinement, effectively suppressing the angular dependence of the transmission dip and mitigating the redshift phenomenon. To evaluate the angular stability of the designed structure, the transmission spectra were simulated under incident angles of 0°, 30°, and 60°, as shown in Fig. S1(c). The results demonstrate that the transmission dip remains nearly invariant across these incident angles, confirming that the proposed metamaterial-based multilayer structure successfully eliminates the angular redshift effect. This robustness against angular variations highlights the potential of the designed structure in practical optical filtering and sensing applications.

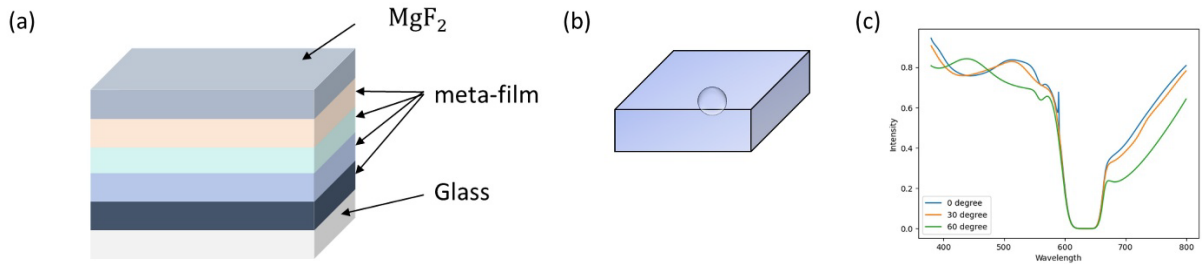

Figure S1. (a) Schematic illustration of the multilayer meta-film structure consisting of alternating material layers on a glass substrate, with an $MgF_2$ ARC layer on top. (b) Schematic illustration of the unit cell of the meta-film. A silver nanoparticle is embedded inside a dielectric layer. (c) Transmission spectra of the meta-film under incident light at different angles (0°, 30°, and 60°), demonstrating the angular independence of spectral response.

**Section 3: Trian and Validation Loss**

As shown in Fig. S2, the training and validation loss curves are closely aligned, indicating that our models do not suffer from overfitting or underfitting issues. Although minor fluctuations were observed during the training process, the losses ultimately stabilized, demonstrating convergence. For the CVAE model, the hyperparameters were set as follows: $\alpha_1 = 2000$, $\beta_1 = 1$, $\gamma_1 = 1$, and $\delta_1 = 1$. For both the forward model and the tandem network, the hyperparameters were set to $\alpha_2 = 300$, $\beta_2 = 1$, and $\gamma_2 = 1$. Upon completion of training, the final test loss for the CVAE model was approximately 3.11, with a reconstruction loss of 0.0005, KLD loss of 1.78, peak wavelength loss of $2.13 \times 10^{-5}$, and peak intensity loss of 0.0013. The forward model achieved a test loss of approximately 1.75, comprising a reconstruction loss of $5.8 \times 10^{-5}$, peak wavelength loss of 1.74, and peak intensity loss of 0.0002. Similarly, the tandem network achieved a final test loss of approximately 1.57, with a reconstruction loss of $3.1 \times 10^{-5}$, peak wavelength loss of 1.57, and peak intensity loss of 0.0001.

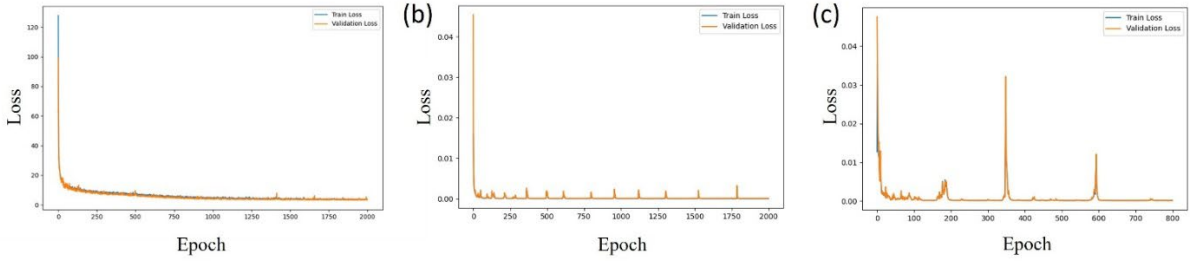

Figure S2. Training and validation loss curves for the single-notch structure: (a) CVAE model; (b) forward model; (c) inverse model.

Meanwhile, we conducted training on the bimodal dataset, and the corresponding training and validation loss curves are presented in Fig. S3. The hyperparameters remain consistent with those used for the single-peak case. Fig. S3(a) illustrates the training and validation loss curves of the CVAE. The test reconstruction loss of the CVAE is 0.00067, the Kullback-Leibler divergence (KLD) loss is 2.3, the peak wavelength loss is 0.0055, and the peak intensity loss is 0.0015. Fig. S3(b) displays the training and validation loss curves of the forward model in the tandem network. The test reconstruction loss for this model is $4.45 \times 10^{-5}$, the peak wavelength loss is 0.006, and the peak intensity loss is $1.63 \times 10^{-5}$. Lastly, the training and validation loss curves of the tandem network are shown in Fig. S3(c). The test reconstruction loss for the tandem network is 0.002, with a peak wavelength loss of 0.13 and a peak intensity loss of 0.002.

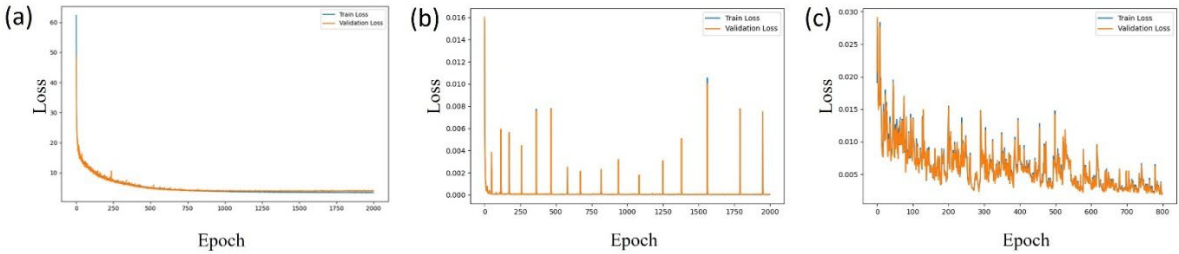

Figure S3. Training and validation loss curves for the double-notch structure: (a) CVAE model; (b) forward model; (c) inverse model.